\documentclass[conference,a4paper]{APSIPA2021}
\usepackage{multirow}
\usepackage[dvips]{graphicx}
\usepackage{amsmath}
\usepackage[psamsfonts]{amssymb}
\usepackage{amsxtra}
\usepackage{threeparttable}
\usepackage[export]{adjustbox}
\usepackage{multicol}
\usepackage{cite}
\usepackage{booktabs}
\usepackage{xcolor}

\begin{document}

\title{An Investigation of Enhancing CTC Model for\\ Triggered Attention-based Streaming ASR}

\author{%
\authorblockN{%
Huaibo Zhao\authorrefmark{1}, 
Yosuke Higuchi\authorrefmark{1}, 
Tetsuji Ogawa\authorrefmark{1}, 
Tetsunori Kobayashi\authorrefmark{1}
}
\authorblockA{%
\authorrefmark{1}
Department of Communications and Computer Engineering, Waseda University, Tokyo, Japan \\
E-mail: zhao@pcl.cs.waseda.ac.jp}
%
}

\maketitle
\thispagestyle{empty}

\begin{abstract}
In the present paper, an attempt is made to combine Mask-CTC and the triggered attention mechanism to construct a streaming end-to-end automatic speech recognition (ASR) system that provides high performance with low latency. The triggered attention mechanism, which performs autoregressive decoding triggered by the CTC spike, has shown to be effective in streaming ASR. However, in order to maintain high accuracy of alignment estimation based on CTC outputs, which is the key to its performance, it is inevitable that decoding should be performed with some future information input (i.e., with higher latency). It should be noted that in streaming ASR, it is desirable to be able to achieve high recognition accuracy while keeping the latency low. Therefore, the present study aims to achieve highly accurate streaming ASR with low latency by introducing Mask-CTC, which is capable of learning feature representations that anticipate future information (i.e., that can consider long-term contexts), to the encoder pre-training. Experimental comparisons conducted using WSJ data demonstrate that the proposed method achieves higher accuracy with lower latency than the conventional triggered attention-based streaming ASR system.
\end{abstract}

\section{Introduction}

The adoption of deep neural networks in automatic speech recognition (ASR) has been proven successful and gained much attention in the past decades~\cite{hinton2012deep, Chorowski2015AttentionBasedMF}. 
Recently, the end-to-end ASR system has shown great simplicity
as a single sequence-to-sequence model
that integrates separated system components
and can be jointly trained with paired audio and text sequences~\cite{Graves2014TowardsES,Chorowski2015AttentionBasedMF,Chan2016ListenAA}.
Attention-based encoder-decoder architecture~\cite{Chorowski2015AttentionBasedMF,Chan2016ListenAA} is one of the major types of architectures for implementation, 
which contains an encoder module that extracts features
and a decoder module that maps the encoder output to the text sequence
using the attention mechanism.
Such architecture could be implemented with various types of neural networks,
including long short-term memory (LSTM) networks~\cite{Hochreiter1997LongSM,Chan2016ListenAA,Moritz2019UnidirectionalNN}
and Transformer~\cite{vaswani2017attention,Dong2018SpeechTransformerAN,Karita2019ImprovingTE}.
Connectionist temporal classification (CTC) is another choice for realizing end-to-end ASR, 
which generates monotonic alignments between the input and output sequences with Markov assumption and dynamic programming~\cite{Graves2014TowardsES}. 
The combined CTC-attention model is often trained to 
further improve the model training and inference processes~\cite{Karita2019ImprovingTE,Karita2019ACS,Chiu2018StateoftheArtSR}.

Streaming property in ASR systems refers to the capability of processing input audio in a real-time fashion
and computing the outputs synchronously, 
which is desired for wide applications in real-world scenarios. 
However, due to the requirements of the global attention mechanism,
the streaming property cannot be easily realized in conventional end-to-end architectures.
Several efforts have been made to introduce the streaming property to the end-to-end ASR system.
Monotonic chunk-wise attention (MoChA)~\cite{Chiu2018MonotonicCA} utilizes soft attention on adaptively chunked subsequences
to permit the attention-based model for streaming processing.
Neural transducer guided by attention-based encoder-decoder is studied in~\cite{Sainath2018ImprovingTP}
to prevent degradation of streaming ASR performance.
In the work of~\cite{moritz2020streaming},
the streaming property is achieved with the triggered attention mechanism,
which generates alignments based on a CTC-based encoder network 
and utilizes the CTC spikes to instruct the trigger-based decoder. 
While the triggered attention-based model enables frame-synchronous decoding
in the attention-based decoder,
its performance highly depends on the accuracy of alignments estimated by the CTC network.
To obtain consistently high recognition accuracy, 
it requires higher latency for the encoder network to generate more accurate CTC alignments, 
which is in opposition to our goal of lowering latency in streaming ASR tasks. Hence, studies on streaming ASR to achieve lower latency while maintaining high accuracy are demanded. 
Similar research has been conducted on the MoChA-based streaming ASR system~\cite{Chiu2018MonotonicCA}  utilizing external hard alignments to minimize latency in~\cite{Inaguma2020MinimumLT}.

In this study, 
we focus on improving the triggered attention-based streaming end-to-end ASR system. 
To this end, we propose to make use of Mask-CTC for enhancing the CTC-based encoder network. 
In Mask-CTC~\cite{higuchi2020mask, higuchi2021improved}, 
an encoder-decoder model is trained with the joint CTC and conditional masked language model (CMLM) objectives~\cite{Devlin2019BERTPO, ghazvininejad2019mask}, 
where the CMLM decoder provides the CTC-based encoder network with the contextual information to improve the performance of CTC. 
Such ability of Mask-CTC to consider long-term context, including information that may be observed in the future, is our key to reliable use of CTC in streaming ASR.
By transferring the pre-trained CTC-based encoder network from the Mask-CTC model, 
we expect to acquire suitable feature representation for streaming objective to reduce the latency requirements of the triggered attention-based streaming ASR as well as improve its performance. 
A three-step training scheme is proposed for efficient model transferring as follows: 
1) learning feature representations that can consider long-term contextual information via Mask-CTC framework; 
2) obtaining a pre-trained model for developing a low-latency streaming ASR model; and 
3) training a triggered attention-based streaming model, where the encoder and CTC modules are initialized with those from 2). 


The rest of the present paper is organized as follows.
Section \ref{sec:techniques} reviews relevant techniques required in our proposed method.
Section \ref{sec:proposal} explains the proposed training strategy for improving triggered attention-based streaming ASR.
Section \ref{sec:exp} describes the experimental setups and evaluation results of our investigation.
Section \ref{sec:conclusion} concludes this paper.



\section{Required Techniques}
\label{sec:techniques}

End-to-end ASR aims to formulate the mapping between input sequence 
$X=(\textbf{x}_t\in\mathbb{R}^D|t=1,...,T )$
and output sequence
$Y=(y_l\in\mathcal{V}|l=1,...,L )$.
Here, $X$ is a $T$-length sequence of speech features, 
where $\textbf{x}_t$ represents the $D$-dimensional feature vector at the $t$-th frame.
$Y$ is an $L$-length sequence of output tokens,
where $y_l$ represents the token at the $l$-th position in the vocabulary $\mathcal{V}$. 

The streaming ASR model that we focus on in this study consists of an attention span-based streaming encoder module~\cite{sukhbaatar2019adaptive} and a triggered attention-based streaming decoder module~\cite{Moritz2019TriggeredAF}.
The CTC-based encoder network obtained from the Mask-CTC framework 
is also utilized for enhancing the streaming model. 
The remainder of this section describes these components.

\subsection{Attention span-based streaming encoder }
\label{ssec:tech-attention-span}

Following the conventional Transformer-based ASR system~\cite{Dong2018SpeechTransformerAN, Karita2019ImprovingTE}, 
our encoder module contains two convolutional neural network (CNN) layers and an $N$ stack of self-attention layers.
Global attention requirement in the self-attention layers prevents the ASR model from performing stream processing. 
To remedy this issue,
we referred to the work of~\cite{sukhbaatar2019adaptive,Chang2020EndtoEndAW}, 
where attention spans are utilized to control the past and future contexts
that the self-attention mechanism attends to.
The conventional self-attention layer calculates attention scores using the scaled dot-product.
Here, we add a span mask $W$ with a fixed size to the scaled dot-product result before softmax operation as: 
\begin{equation}
    \mathrm{Attention}(Q,K,V) =
    \mathrm{softmax} \bigg(
    \frac{Q{K^T}}{\sqrt{d_k}}W
    \bigg) V,
\end{equation}
where $Q\in\mathbb{R}^{n_q\times d_q}$,
$K\in\mathbb{R}^{n_k\times d_k}$,
and $V\in\mathbb{R}^{n_v\times d_v}$ denote query,
key, and value matrices respectively, with $n_*$ as sequence length and $d_*$ as feature dimension.
In our work,
since a controllable latency of the system is aimed for,
restrictions are only set on the range of the future contexts,
while attending to the past contexts is always allowed.

\subsection{Triggered attention-based streaming decoder}
\label{ssec:tech-tasd}

For introducing the streaming property into the decoder module,
we adopted the triggered attention mechanism~\cite{Moritz2019TriggeredAF, moritz2020streaming},
which consists of an encoder for feature extraction, a CTC module (trigger network) for trigger generation, and an attention-based decoder.
The encoder transforms an input feature sequence into an encoded feature sequence.
The CTC module computes spike-like posteriors (i.e., triggers).
Using such spike-like posteriors,
the pause-delimited speech utterances can be dynamically partitioned into smaller subsequences according to the input.
At the starting point of this subsequence,
the attention mechanism is triggered,
and the encoded feature frames preceded by this trigger event and some look-ahead frames are decoded~\cite{Moritz2019TriggeredAF}.
This framework can achieve frame-synchronous decoding.

The triggered attention-based decoder requires the alignment between the encoded feature sequence $H$ and output label sequence $Y$,
which can be computed based on the CTC outputs and utilized for instructing the training of the decoder.
In practice,
alignment information is provided
by Viterbi decoding with a CTC-based trigger network
and used so that the attention-based decoder could be conditioned solely on the past encoded frames and past outputs as:
\begin{equation}
    P_{\text{ta}}(Y|H) = \prod_{l=1}^{L} P(y_l|y_{1:l-1}, \textbf{h}_{1:{n_l}}),
\end{equation}
where $n_l$ denotes the first occurring position of the symbol $y_l$ in the CTC alignment.
This formulation, which does not explicitly need feature extraction on future, is suitable for the purpose of online ASR.

During training,
all modules are initialized with a pre-trained non-streaming Transformer model and fine-tuned for the streaming purpose;
the encoder module is trained with attention spans as described in Sect.~\ref{ssec:tech-attention-span} .
During run-time,
joint CTC-attention decoding is carried out.

\subsection{Mask-CTC}
\label{ssec:tech-maskCTC}

Mask-CTC~\cite{higuchi2020mask} is a framework for learning feature representations
to achieve high accuracy and fast inference speed, 
and is the cornerstone of this proposal.
Following the encoder-decoder architecture, the Mask-CTC framework tackles the speech recognition task with a two-pass approach. In the first pass, CTC output is utilized to generate initial inference output from the encoded speech feature sequence, where less confident tokens are masked and refined by mask prediction conducted by the Transformer decoder in the second pass. During training, part of the ground truth is randomly substituted with a mask token to train the Transformer decoder as a conditional masked language model (CMLM) to resolve masked tokens \(Y_{\text{mask}}\) based on observed tokens \(Y_{\text{obs}}\) and encoder sequence \(X\). 
The CMLM decoder is modeled as follows: 
\begin{equation}
    P_{\text{cmlm}}(Y_{\text{mask}}|Y_{\text{obs}}, X)  = \prod\limits_{y\in Y_{\text{mask}}} P_{\text{cmlm}}(y|Y_{\text{obs}}, X).
\end{equation}
Through the training based on the CTC and CMLM objectives~\cite{Devlin2019BERTPO, ghazvininejad2019mask}, its CTC model is enhanced with the contextual information captured in the decoder and achieves competitive output accuracy to autoregressive decoding results. As when we only apply CTC prediction, it significantly outperforms the greedy CTC prediction in hybrid CTC-attention models and achieves competitive results to autoregressive Transformer decoding. The encoder model is enhanced to perform better feature extraction on future contexts. Meanwhile, a better alignment between the encoded feature sequence and the output label sequence is achieved by the CTC model. 
\begin{figure*}
\sbox0{\includegraphics{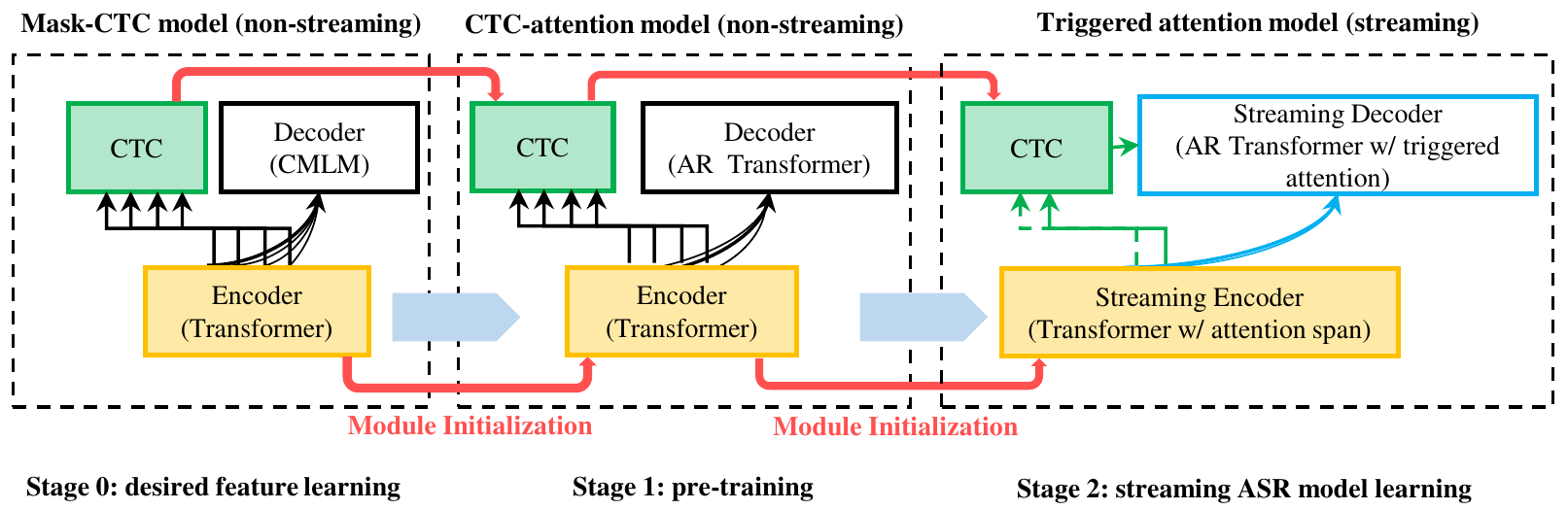}}
\centering
\usebox0
\caption{Procedure of building enhanced triggered attention-based streaming model
for low latency and high accuracy streaming ASR:
Stage 0 (left) for anticipatory feature learning, Stage 1 (center) for pre-training, and Stage 2 (right) for streaming model learning.
In Stage 0,
non-streaming Mask-CTC model is trained with CTC-CMLM objective to obtain a desirable (i.e., capable of handling long-term contexts) feature extraction mechanism;
in Stage 1, non-streaming CTC-attention model for autoregressive Transformer-based decoding is initialized with Mask-CTC modules for better feature representations and trained with CTC-attention objective;
in Stage 2, enhanced CTC-attention model pre-trained in Stage 1 is used to generate alignments and initialize encoder and CTC modules for learning streaming model with triggered attention-based decoder and attention span-based streaming encoder.}
\label{fig:proposal}
\end{figure*}

\section{Feature learning for low latency and high accuracy streaming ASR}
\label{sec:proposal}

For a study of streaming property in ASR systems,
we focus on triggered attention-based streaming ASR in~\cite{moritz2020streaming},
which generates alignment information with the CTC-based encoder network and performs autoregressive decoding triggered by the CTC spikes.
It, however, should be noted that
higher latency is required to maintain consistently high recognition accuracy in this framework,
since knowing the future context contributes to reliable alignment and accurate recognition.
We, therefore, attempt to enhance the triggered attention-based streaming model
by acquiring feature representations suitable for low latency and high accuracy streaming ASR.
For that purpose, 
our proposal takes advantage of Mask-CTC's ability to provide better feature extraction for the streaming ASR model by considering long-term contexts including the future. 

The remainder of this section gives an overview (Sect. \ref{ssec:proposal-overview}) and details (Sect. \ref{ssec:proposal-training}) of the proposed procedure for building the streaming ASR model,
as well as the aim of pre-training the streaming ASR model using Mask-CTC (Sect. \ref{ssec:proposal-transformer}),
which is the key to our proposal.



\subsection{Overview of building streaming model}
\label{ssec:proposal-overview}

To achieve low latency and high accuracy streaming ASR, 
the enhanced triggered attention-based model is constructed
by the following three-step process as shown in Fig.~\ref{fig:proposal}:
\begin{itemize}
\item \textbf{Stage 0 (Anticipatory feature learning)} aims to learn 
the process of extracting feature representations considering long-term contextual information using the Mask-CTC framework (with the CTC-CMLM objective).

\item \textbf{Stage 1 (Pre-training for reliable streaming ASR)}
aims to obtain a pre-trained model for developing low-latency streaming ASR models efficiently in the latter stage.
In this stage,
following the pre-training process in the triggered attention mechanism,
a CTC-attention model for autoregressive Transformer-based decoding 
is built; it differs from the existing approach in that the encoder and CTC modules are initialized with those of Mask-CTC 
for better feature representations.

\item \textbf{Stage 2 (Streaming ASR model learning)}
aims to construct the triggered attention-based streaming model using the reliable alignment information 
and initial parameters,
all of which are transferred from the pre-trained CTC-attention model built in Stage 1.
\end{itemize}

Note that
the effects of Mask-CTC are used directly in the pre-training stage (Stage 1)
and propagated implicitly to the target streaming ASR model in its building stage (Stage 2). 
Efforts have been made to directly initialize Mask-CTC components in the construction of target streaming ASR model (i.e., skipping Stage 1). However, due to the lack of suitability for autoregressive decoding, the streaming ASR model showed worse performance than the conventional model. 


\subsection{Effectiveness of using Mask-CTC for feature learning}
\label{ssec:proposal-transformer}

The enhancement of the encoder for streaming is inspired by the high recognition accuracy of the greedy CTC outputs of Mask-CTC~\cite{higuchi2020mask}.
The superiority of Mask-CTC over the CTC-attention model in terms of accuracy of CTC outputs suggests that
multitask learning of CTC and CMLM objectives provides higher accuracy in the feature extraction process in the encoder and thus higher accuracy in the alignment given by the CTC module.
Since CMLM tries to solve the prediction problem considering the long-term context from the past to the future,
learning encoders based on the CMLM objective is expected to contribute to the learning of the feature extraction process that enables anticipation.
Such properties are desirable in streaming ASR tasks,
especially when only low latency is allowed.

When training a triggered attention-based streaming model,
alignment and pre-training of each module are necessary,
and the non-streaming CTC-attention model has been used for this purpose.
In the proposed method,
such CTC-attention model is initialized with the Mask-CTC modules and trained in Stage 1,
and 
we expect that
the effects of Mask-CTC, which is suitable for learning feature representations for low latency and high accuracy streaming, will be introduced into this pre-trained model.


\subsection{Detailed procedure of training streaming ASR model}
\label{ssec:proposal-training}

In Stage 1,
a CTC-attention model consisting of a Transformer encoder module, a CTC module, and an autoregressive Transformer decoder module is pre-trained for the triggered attention-based streaming model.
In the proposed method,
the encoder and CTC modules are initialized
with the corresponding modules of Mask-CTC obtained in Stage 0,
and these modules with enhanced feature extraction process as well as the autoregressive Transformer decoder are then trained with the hybrid CTC-attention objective.
The Transformer encoder enhanced by Mask-CTC is expected to yield more accurate CTC outputs than the Transformer encoder in the conventional CTC-attention model.
Since Mask-CTC is designed for non-autoregressive and non-streaming ASR,
this stage remains in the non-streaming manner
and focuses on obtaining the initial values of the encoder and CTC modules that make up the triggered attention-based streaming mechanism,
as well as the reliable alignments needed to train its decoder module. 

In Stage 2,
a triggered attention-based streaming model is constructed.
The procedure for training and operating this model follows Sect. \ref{ssec:tech-tasd},
but it differs from the conventional approach in that
the enhanced CTC-attention model constructed in Stage 1 is used to generate alignment information and serves as a pre-trained model for the streaming model.
Since streaming ASR is designed for real-time audio processing,
high feature representation extraction capability is required even in situations where only low latency is acceptable (i.e., shorter look-ahead frames available).
The pre-trained CTC-attention model incorporates Mask-CTC's superiority in better feature extraction on future.
It, therefore, is expected to generate reliable alignments even with low latency and provide better instructions to the triggered attention mechanism.
In addition,
training a streaming model from such a pre-trained model is expected to provide more accurate recognition than existing streaming ASR models, especially at low latency.


\section{Streaming Speech Recognition Experiments}
\label{sec:exp}

Our main proposal is to utilize the framework of Mask-CTC to acquire feature representations suitable for triggered attention-based streaming ASR with low latency and high accuracy.
To this end,
we conducted a preliminary experiment to validate the adequacy of pre-trained feature representations obtained from Mask-CTC (Experiment 1)
and an experiment to investigate the effectiveness of the developed streaming ASR model (Experiment 2).

\subsection{Datasets}
\label{ssec:exp-data}

In the experiments,
our models were trained and tested using the 81-hour Wall Street Journal (WSJ) corpus~\cite{Paul1992TheDF}, 
which includes English utterances from read news articles. 
As input speech features, 
we used 80-dimensional log-Mel spectral energies plus three extra pitch features using~\cite{povey2011kaldi}.
We used characters (i.e., Latin alphabets) for tokenizing output texts. 
\begin{table*}[t]
\begin{center}
\caption{Effectiveness of Initializing CTC-attention ASR Model with Mask-CTC.
\textbf{Greedy CTC outputs} from non-streaming ASR systems such as conventional CTC-attention, Mask-CTC, and enhanced CTC-attention models were compared in terms of word error rates (\%) on WSJ dataset.}
\label{table1}
\begin{tabular}{ll|l|l|c|c} 
    \toprule
    \textbf{Model} & & \textbf{Encoder} & \textbf{Decoder} & \textbf{eval92} & \textbf{dev93}\\
    \midrule
    CTC-attention & (Conventional) & Transformer randomly initialized & AR Transformer & 16.6 & 21.2\\
    Mask-CTC & (Stage 0 in Fig.~\ref{fig:proposal}) & Transformer randomly initialized & CMLM Transformer & 13.9 & 17.9\\
    Enhanced CTC-attention & (Stage 1 in Fig.~\ref{fig:proposal}) & Transformer pretrained with Mask-CTC & AR Transformer & \bf{13.3} & \bf{16.8}\\
 \bottomrule
\end{tabular}
\end{center}
\end{table*}
\begin{table*}[t]
\begin{center}
\caption{Effectiveness of using Mask-CTC in pre-training of triggered attention-based streaming ASR model. Existing and enhanced streaming ASR models were compared in terms of word error rates (\%) along with encoder latency (ms) on WSJ dataset.}
\label{table2}
\begin{tabular}{l|l|c|c|c} 
    \toprule
    \textbf{Model} & \textbf{Pre-trained model} & \textbf{Encoder latency} [ms] & \textbf{eval92} & \textbf{dev93}\\
    \midrule
    \multirow{4}{*}[0pt]{Triggered attention-based streaming ASR } & \multirow{4}{*}[0pt]{CTC-attention} & 160 & 28.2 & 34.5\\
    & & 320 & 22.4 & 27.5\\
    & & 480 & 18.9 & 24.1\\
    & & 640 & 17.0 & 20.2\\
    \midrule
    \multirow{4}{*}[0pt]{Enhanced triggered attention-based streaming ASR} & \multirow{4}{*}[0pt]{Enhanced CTC-attention} & 160 & 21.3 & 25.9\\
    & & 320 & 15.5 & 19.5\\
    & & 480 & 14.3 & 19.1\\
    & & 640 & 14.1 & 18.1\\
    \bottomrule
\end{tabular}
\end{center}
\end{table*}

\subsection{Experimental setup}
\label{ssec:exp-setup}

All the end-to-end ASR models were constructed using the same Transformer-based encoder-decoder architecture as in~\cite{Karita2019ImprovingTE}, 
where the encoder and decoder modules contained twelve and six self-attention layers, respectively.
Each self-attention layer was implemented with 256 hidden units, 2048 feed-forward inner dimensions, and four attention heads.
For training non-streaming models,
the learning rate was set to 10.0 and the training converged at around 50 through 100 epochs.
For training streaming models,
the learning rate was set to 1.0 and the number of training epochs was set to 120.
The final models were obtained by averaging model parameters from the last five snapshots. 
For decoding with CTC only, 
we used the best path decoding algorithm~\cite{graves2006connectionist}, 
where a greedy output is obtained by suppressing repeated tokens and removing blank symbols. For joint CTC-triggered attention decoding, we utilized the frame-synchronous one-pass decoding algorithm~\cite{moritz2020streaming}, which integrated triggered attention decoding with the frame-synchronous prefix beam search algorithm from~\cite{Maas2014FirstPassLV}. 
The CTC-attention weight for the joint decoding was set to 0.5 and the beam search was conducted with the beam size of 10. 

\subsection{Evaluation items}
\label{ssec:exp-items}

We describe the models that are compared in the two experiments.
All systems were developed using ESPnet~\cite{Watanabe2018ESPnetES}
and compared without the usage of external language models.

\subsubsection{Experiment 1}

First,
we investigated the adequacy of introducing the Mask-CTC's effect into the pre-training stage of the streaming ASR model (Stage 1), 
which aims for better feature extraction on future contexts.
To this end, 
we evaluated the greedy CTC outputs generated from the following models: 
\begin{itemize}

\item \textbf{CTC-attention}:
A conventional non-streaming model, 
consisting of an autoregressive (AR) Transformer-based decoder, 
a Transformer-based encoder, and 
a CTC module.

\item \textbf{Mask-CTC}:
A non-streaming model with a Transformer-based CMLM decoder, a Transformer-based encoder, and a CTC module (see Sect.~\ref{ssec:tech-maskCTC} for details). 
The model is constructed in Stage 0 of the proposed learning method. 

\item \textbf{Enhanced CTC-attention}:
A non-streaming CTC-attention model enhanced by Mask-CTC in Stage 1: 
the encoder and CTC modules are initialized with those of the Mask-CTC model and all modules are trained with the CTC-attention objective. 
\end{itemize}

The comparison of these models was designed to clarify the effectiveness of encoder and CTC modules in Enhanced CTC-attention. 
Here,
if Enhanced CTC-attention achieves best recognition results with the CTC outputs, 
this model can be used to improve the triggered attention-based streaming ASR model, 
being expected to generate more accurate alignment information and give better initialization for the encoder and CTC modules. 

\subsubsection{Experiment 2}

Secondly,
we investigated the effectiveness of the proposed triggered attention-based streaming ASR model, 
which was constructed based on Mask-CTC-based feature learning in Stage 2.
The streaming performances were evaluated in terms of the recognition accuracy and latency.
For this purpose,
we compared the following streaming ASR models:
\begin{itemize}

\item \textbf{Triggered attention-based streaming}:
An existing triggered attention-based streaming ASR model trained with the CTC-attention objective~\cite{moritz2020streaming};
alignment generation and module initialization are conducted using the existing CTC-attention model examined in Experiment 1.

\item \textbf{Enhanced triggered attention-based streaming}:
An enhanced triggered attention-based streaming model, 
whose feature representation learning implicitly utilizes Mask-CTC;
alignment generation and module initialization are conducted using the enhanced CTC-attention model examined in Experiment 1.

\end{itemize}

The joint CTC-attention decoding was performed in both systems.
To analyze the streaming characteristics of the model,
we investigated the impact of encoder latency values on the recognition accuracy.
Here, the encoder latency is defined as the look-ahead range for the encoder to perform feature extraction for each input frame, 
which is controlled by the attention span mechanism as explained in Sect.~\ref{ssec:tech-attention-span}. 
Our goal is to achieve high recognition accuracy while reducing this encoder latency.

\subsection{Experimental results}
\label{ssec:exp-result}

\subsubsection{Experiment 1}
The results of Experiment 1 are summarized in Table~\ref{table1}.
This table lists the word error rates (WERs) of CTC outputs from three models on the WSJ dataset.
The superiority of the Mask-CTC model to the CTC-attention model demonstrated that
the CTC-CMLM objective performs better than the CTC-attention objective
for more accurate CTC outputs.
The enhanced CTC-attention model
yielded the best accuracy.
This result suggests that
feature representation learning based on Mask-CTC can contribute to the learning of CTC-attention models with AR decoders, 
even if the modules are not directly trained on the CTC-CMLM objective. 
It also suggests that the enhanced CTC-attention model is effective for pre-training the triggered attention-based streaming ASR model.

\subsubsection{Experiment 2} Table~\ref{table2} shows the results of Experiment 2.
This table lists the WERs (\%) of the conventional and enhanced triggered attention-based streaming ASR model along with the encoder latency values (ms) computed on the WSJ dataset.

This result demonstrates that
the proposed enhancement of the triggered attention-based streaming ASR model reduced the WERs of the existing model (e.g., 2.1\% to 8.6\%),
regardless of the encoder latency values.

In addition,
the results demonstrate that
the proposed pre-training can achieve low latency processing while maintaining recognition accuracy
in the triggered attention-based streaming ASR.
Table \ref{table2} shows that
for both systems, 
the WERs increased as the encoder latency values decreased.
Note that the degradation in recognition accuracy was caused by the small number of future contexts given to the encoder for feature extraction.
While the latency change had a significant effect on the recognition accuracy in the conventional streaming ASR model,
the performance degradation was moderated when the proposed learning method was applied.
In fact,
when the encoder latency value is reduced from 640 ms to 320 ms,
the enhanced model maintained almost the same accuracy level with around 1.4\% degradation
while the existing model increased WER by 7.3\%.
Furthermore,
the proposed learning method achieved higher recognition accuracy with lower latency than using the existing models.
For example,
the performance of the proposed model with 320 ms latency (e.g., 15.5\% for eval92 and 19.5\% for dev93) outperformed the performance of the existing model with 640 ms latency (e.g., 17.0\% for eval92 and 20.2\% for dev93).
Such higher resilience towards latency reduction indicates that
our proposal contributes to better feature extraction and alignment generation with less future context from the input sequence.
On the other hand,
the proposed method still showed a sharp performance drop
when the latency value decreased from 320 ms to 160 ms,
showing the existence of a bottom line for latency requirements. 

\section{Conclusions}
\label{sec:conclusion}

In the present paper,
we investigated a method to build a low latency and high accuracy streaming end-to-end ASR system.
We attempted to enhance the triggered attention-based model by transferring the property of Mask-CTC,
which provides better feature extraction on future context, for lower latency.
Experimental comparisons demonstrated that
the proposed method yielded lower encoder latency and higher recognition accuracy than the conventional system.
In the future,
we will further study the trade-off between the performance and latency,
and investigate our method on various language datasets to focus on the possibility of wider use.

\renewcommand{\textheight}{105mm}

\bibliographystyle{IEEEbib}
\bibliography{refs}

\end{document}